\documentclass[12pt]{article}
\usepackage[english,german,french,polish]{babel}
\usepackage[T1]{fontenc}
\usepackage{amsfonts}

\begin{document}

\selectlanguage{english}

\baselineskip 0.76cm
\topmargin -0.6in
\oddsidemargin -0.1in

\let\ni=\noindent

\renewcommand{\thefootnote}{\fnsymbol{footnote}}

\newcommand{\SM}{Standard Model }

\pagestyle {plain}

\setcounter{page}{1}

\pagestyle{empty}

~~~

\begin{flushright}
IFT-- 05/21
\end{flushright}

\vspace{1.0cm}

{\large\centerline{\bf Predicting the tauon mass as well as the ratio}} 

{\large\centerline{\bf of electron and muon masses}} 

\vspace{0.4cm}

{\centerline {\sc Wojciech Kr\'{o}likowski}}

\vspace{0.3cm}

{\centerline {\it Institute of Theoretical Physics, Warsaw University}}

{\centerline {\it Ho\.{z}a 69,~~PL--00--681 Warszawa, ~Poland}}

\vspace{0.6cm}

{\centerline{\bf Abstract}}

\vspace{0.2cm}

Treating the Koide equation and another efficient charged-lepton mass formula (having 
the form of a mass sum rule) as a system of two mathematically independent algebraic 
equations for three charged-lepton masses, we predict the tauon mass as well as the ratio 
of electron and muon masses, both consistent with experiment.

\vspace{1.0cm}

\ni PACS numbers: 12.90.+b 

\vspace{1.0cm}

\ni September 2005  

\vfill\eject

~~~
\pagestyle {plain}

\setcounter{page}{1}

\vspace{0.2cm}

\ni {\bf 1. Two mass formulae}

\vspace{0.2cm}

In particle physics literature, two essentially empirical formulae are published, predicting very well the tauon mass (experimentally measured as $m_\tau = 1776.99^{+0.29}_{-0.26}\;{\rm MeV}$) in terms of the electron and muon masses (experimentally equal to $m_e = 0.5109989$ MeV and $m_\mu = 105.65837$ MeV) [1].

The first of them is the implicit and nonlinear mass equation discovered by Koide in 1981 [2]:

%rownanie 1
\begin{equation}
m_e + m_\mu + m_\tau = \frac{2}{3} (\sqrt{m_e} + \sqrt{m_\mu} + \sqrt{m_\tau})^2 \,.
\end{equation}

\ni It gets two solutions:

%rownanie 2\pm
$$
\;\;\;\;m_\tau  =  \left[ 2\left( \sqrt{m_e}+ \sqrt{m_\mu} \right) \pm \sqrt{3(m_e + m_\mu) + 12\sqrt{m_e m_\mu}}\, \right]^2  \eqno(2\pm)
$$

\vspace{-0.5cm} 

%rownanie 3\pm
$$
 = \; \left\{\begin{array}{l} 1776.9689\; {\rm MeV}\\ 3.3173557\; {\rm MeV} \end{array} \right.\; = \;\left\{\begin{array}{l} 1776.97\; {\rm MeV}\\ 3.31736\; {\rm MeV} \end{array} \right.  \;\;, \eqno(3\pm)
$$

\addtocounter{equation}{+2}

\ni where -- in the second step -- the experimental values of $m_e$ and $m_\mu $ are used as an input, leading to two different predictions for $m_\tau $. The large prediction is excellent, almost equal to the central value of experimental $m_\tau $, while the small one does not correspond to any known experimental object. The small solution implying this small prediction cannot be excluded from the Koide  equation (1) by itself, unless it is additionally required that $m_\mu <m_\tau $ (for the experimental $m_e $ and $m_\mu $ as an input).

The second of the mentioned mass formulae, found out in 1992 [3], is the explicit and linear mass sum rule:

%rownanie 4,5
\begin{eqnarray}
m_\tau & = & \frac{6}{125}(351 m_\mu - 136 m_e) \\ & = &1776.7964\;{\rm MeV} = 1776.80\;{\rm MeV}\,,
\end{eqnarray}

\ni where -- in the second step -- the experimental values of $m_e$ and $m_\mu $ are inserted as an input, giving a prediction also consistent with the experimental $m_\tau $. This mass sum rule follows exactly from the mass spectral formula conjectured for the charged leptons in 1992 [3]:

%rownanie 6
\begin{equation}
m_i = \mu\,  \rho_i  \left(N^2_i + \frac{\varepsilon  - 1}{N^2_i} \right) \;\;(i = e, \mu, \tau) \,,
\end{equation}

\ni where $N_i $ and $\rho_i $ denote

%rownanie 7
\begin{equation}
N_i = 1,3,5 \;,\;\; \rho_i = \frac{1}{29} \,,\,\frac{4}{29} \,,\,\frac{24}{29} 
\end{equation}

\ni ($\sum_i \rho_i = 1$), while $\mu  > 0$ and $\varepsilon  > 0$ are two constants whose values can be determined (simultaneously with $m_\tau $) by the input of experimental $m_e$ and $m_\mu $:

%rownanie 8
\begin{eqnarray}
\mu  & = & \frac{29(9m_\mu - 4m_e)}{320} = 85.992356\;{\rm MeV}\;,\nonumber \\ 
\varepsilon & = & \frac{320 m_e}{9m_\mu - 4m_e} = 0.1723289 \,.
\end{eqnarray}
  
\ni In fact, these values follow from the first and second of three equations:

%rownanie 9
\begin{equation}
m_e = \frac{\mu }{29} \varepsilon  \;, \nonumber \\  m_\mu = \frac{\mu }{29} \frac{4}{9} (80 +  \varepsilon ) \;,\;  m_\tau = \frac{\mu }{29} \frac{24}{25} (624 + \varepsilon )\;,
\end{equation}

\ni presenting explicitly the formula (6). Then, from the third equation, the mass sum rule (4) follows.

Although essentially empirical, Eq. (6) is supported by a speculative background [3], where the numbers (7) get simple interpretations. In particular, $N_i -1 = 0,2,4$, is the number of additional algebraic spin-1/2 partons appearing in three fermion generations. These additional algebraic  partons obey the Pauli principle, causing the existence of three and only three fermion generations.

In a recent note [4], it has been pointed out that {\it numerically}, for the experimental $m_e$ and $m_\mu $ as well as the predicted $m_\tau $ given in Eq. (5), the formula (4) is an excellent approximate solution to the Koide equation (1) and {\it vice versa}. In fact, for these values of $m_e$, $m_\mu $ and $m_\tau $ one obtains

%rownanie 10
\begin{equation}
m_e + m_\mu + m_\tau = \frac{2}{3}\frac{1}{1.0000146}(\sqrt{m_e} + \sqrt{m_\mu} + \sqrt{m_\tau})^2  \;.
\end{equation}

\ni Of course, this equation is a way of expressing that both mass formulae are numerically very similar.

\vspace{0.2cm}

\ni {\bf 2. How to predict the ratio of $m_e$ and $m_\mu $?}

\vspace{0.2cm}

In this note, we would like to present another observation. As is not difficult to see, the right hand sides of Koide equation (2+) for $m_\tau $ and of formula (4) also for $m_\tau $, treated as functions of two variables $m_e$ and $m_\mu $, are mathematically independent. The difference of these two functions is equal to 0.1725 MeV at the point of experimental $m_e$ and $m_\mu $, and so is contained in the limits of experimental uncertainty of $m_\tau $. Thus, up to this difference of the order $10^{-2}\%$, the functions (2+) and (4) express {\it the same variable} $m_\tau $ through $m_e$ and $m_\mu $ {\it in two independent ways}: they form a system of two mathematically independent, homogeneous equations for three variables $m_e $, $m_\mu $ and $m_\tau $. Hence, eliminating $m_\tau $, we get the following equation for the ratio $ r \equiv m_\mu /m_e $:

%rownanie 11
\begin{equation}
\left[2(\sqrt{r} +1) + \sqrt{3(r+1)} + 12\sqrt{r} \right]^2 - \frac{6}{125}(351r -136) = 0 \,.
\end{equation}

\ni Making use of the Mathematica5 program, we solve Eq. (11) immediately, obtaining the unique solution 

%rownanie 12
\begin{equation}
r = 206.99287 
\end{equation}

\ni which agrees not so badly with the experimental $r = 206.76829$. Here, the error of the order $10^{-1}\%$ is a consequence of the difference 0.1725 MeV of the order $10^{-2}\%$ between the two $m_\tau $ values (3+) and (5), both contained in the limits of experimental uncertainty of $m_\tau $ (the same is true for their difference).

In order to verify the selfconsistency of our procedure, one may try to take into account the difference 0.1725 MeV between the Koide value (3+) of $m_\tau $ and its value (5). When divided by the experimental $m_e $, this difference (considered as 0.1725000 MeV) leads to the number $0.3375741 \simeq 1/3$. If one puts such a number in place of 0 on the right hand side of Eq. (11) one gets the "corrected" equation for the ratio $ r \equiv m_\mu /m_e $:

%rownanie 13
\begin{equation}
\left[2(\sqrt{r} +1) + \sqrt{3(r+1)} + 12\sqrt{r} \right]^2 - \frac{6}{125}(351r -136) = 0.3375741 \,.
\end{equation}

\ni Then, the Mathematica5 program gives uniquely the solution

%rownanie 14
\begin{equation}
r = 206.76831 \,.
\end{equation}

\ni that coincides with the experimental $r = 206.76829$, showing the selfconsistency of our procedure.

\vspace{0.2cm}

\ni {\bf 3. Some improvements}

\vspace{0.2cm}

From this exercise, we can also see that a {\it theoretical} correction to the mass sum rule (4) -- which could make the small difference 0.1725 MeV  between the Koide value (3+) and the value (5) still smaller -- would improve Eq. (11) and so, its solution (12). Such an improvement may be achieved, for instance, by introducing into the diagonal charged-lepton mass matrix presented in Eq. (6) some nondiagonal matrix elements as {\it e.g.} those proposed in Ref. [5]. Then, however, a third free parameter is introduced into the formula (6) (beside two constants $\mu $ and $\varepsilon $).

In the present note, we propose to correct the formula (6) in another way, without introducing nondiagonal matrix elements and a third free parameter. To this end, in place of Eq. (6) we consider a corrected mass spectral formula of the form

%rownanie 15
\begin{equation}
m_i = \mu \, \rho_i  \left(N^2_i + \frac{\varepsilon(1+ \delta_i)  - 1}{N^2_i} \right) \;\;(i = e, \mu, \tau) \,,
\end{equation}

\ni where we assume that the correction $\delta_i$ appears only for $i = \tau $, and is given specifically as

%rownanie 16
\begin{equation} 
\delta_i =\frac{(N_i-1)(N_i -3)}{N_i^2} = \frac{8}{25}\delta_{i \tau} = 0.32 \delta_{i \tau}
\end{equation}

\ni (due to $N_i = 1,3,5$). The corrected mass spectral formula (15) can be rewritten in the explicit form

%rownanie 17
\begin{equation}
m_e = \frac{\mu }{29} \varepsilon  \;, \nonumber \\  m_\mu = \frac{\mu }{29} \frac{4}{9} (80 +  \varepsilon ) \;,\;  m_\tau = \frac{\mu }{29} \frac{24}{25} (624 + \varepsilon ) + \delta m_\tau\;,
\end{equation}

\ni where (due to the relation $\mu\,\varepsilon = 29 m_e$) the correction $\delta m_\tau $ is

%rownanie 18
\begin{equation}
\delta m_\tau = \frac{\mu }{29} \frac{24}{25} \varepsilon \delta_\tau = \frac{6}{125} \frac{32}{5} m_e = 0.3072 m_e
\end{equation}

\vspace{-0.5cm}

%rownanie 19
\begin{equation}
 = 0.1569789\;{\rm MeV}\;,
\end{equation}

\ni when -- in the second step -- the experimental $m_e$ is used. Then

%rownanie 20
\begin{equation}
m_\tau = (1776.7964 + 0.1570)\,{\rm MeV} = 1776.9534\,{\rm MeV}
\end{equation}

\ni with the use of experimental $m_e$ and $m_\mu $ (and so, the experimental $\mu $ and $\varepsilon $). Here, $\mu $ and $\varepsilon $ are related to $m_e$ and $m_\mu $ still through Eqs. (8). The prediction (20) for $m_\tau$ is close to the central value of experimental $m_\tau = 1776.99^{+0.29}_{-0.26}$, one order of magnitude closer than the previous prediction (5) (also consistent with experiment). It is easy to see that now the mass sum rule (4) is corrected to the form

%rownanie 21,22
\begin{eqnarray}
m_\tau & = & \frac{6}{125}\left[351 m_\mu - \left(136  - \frac{32}{5}\right) m_e\right]  = \frac{162}{625}(65 m_\mu - 24 m_e) \\ & = & 1776.9534\,{\rm MeV} = 1776.95\;{\rm MeV} \,,
\end{eqnarray}

\ni when -- in the second step -- the experimental $m_e$ and $m_\mu $ are applied.

The correction $\delta_i $ in the mass spectral formula (15), appearing only for the third generation $i = \tau $, may be connected with the speculative structure of three fermion generations [3]. In the case of this structure, in a fermion belonging to the third generation there are two pairs of additional algebraic spin-1/2 partons ($N_\tau -1 = 2\cdot 2 = 4$), while in a fermion of the second generation there is only one such pair ($N_\mu -1 = 1\cdot 2 = 2$) and in a fermion of the first generation  -- none ($N_e -1 = 0\cdot 2 = 0$). Such two pairs existing simultaneously may develop a new intrinsic mutual interaction manifesting itself as a correction $\delta_i = \delta_\tau \delta_{i \tau}$ in the new mass spectral formula (15).

Now we repeat essentially our previous argument leading to the equation (11) for the ratio $r \equiv m_\mu /m_e $. We neglect the difference 0.0155 MeV of the order $10^{-3}\%$ between the Koide solution (2+) for $m_\tau $ and the new formula (21) also for $m_\tau $ at the point of experimental $m_e$ and $m_\mu $, and then, eliminating $m_\tau $ from such a system of two mathematically independent, homogeneous equations for $m_e$ and $m_\mu $, we obtain in place of Eq. (11) the corrected equation for the ratio $r \equiv m_\mu /m_e $:

%rownanie 23
\begin{equation}
\left[2(\sqrt{r} +1) + \sqrt{3(r+1)} + 12\sqrt{r} \right]^2 - \frac{6}{125}\left(351r -136+ \frac{32}{5}\right) = 0 \,.
\end{equation}

\ni In this case, the Mathematica5 program leads to the unique solution

%rownanie 24
\begin{equation}
r = 206. 78852 
\end{equation}

\ni that is close to the experimental $r = 206. 76829$, one order of magnitude closer than the solution (12) of equation (11). Now, the error of $r$ is of the order $10^{-2}\%$.

\vspace{0.2cm}

\ni {\bf 4. Conclusions}

\vspace{0.2cm}

The result presented in this note is by no means a surprise in the specific situation, when there are two mathematically independent equations for the same three variables $m_e$, $m_\mu $ and $m_\tau $, giving the same value of $m_\tau $ at the point of experimental $m_e$ and $m_\mu $ (up to a small difference contained in the limits of experimental uncertainty). However, a real surprise is the {\it simultaneous validity} of two such independent equations: the Koide formula (1) (implying the solution (2+)) and the mass sum rule (4) or, better, (21).

A fargoing conclusion we can come to in the present note is that the simultaneous validity of the Koide equation (1) and the mass sum rule (4) or, better, (21) predicts the ratio $r \equiv m_\mu /m_e $, leaving, say, $m_e$ as an only experimental parameter. Then, this parameter, put equal to the experimental $m_e$, predicts two remaining charged-lepton masses $m_\mu $ and $m_\tau $, consistently with their experimental values. However, the errors of so calculated $m_\mu $ and $m_\tau $ are dependent on the error of $r$ evaluated from Eq. (11) or (23) that is of the order $10^{-1}\%$ or $10^{-2}\%$ , respectively (and follows from a difference of the order $10^{-2}\%$ or $10^{-3}\%$ between two $m_\tau $ values (3+) and (5) or (22)). In contrast, $m_\tau $ calculated from Eq. (2+) and also Eq. (4) or (21) at the point of experimental $m_e$ and $m_\mu $ is not burdened with the error of $r$. Using the better value (24) of $r$ evaluated from Eq. (23), as well as the experimental $m_e$, we predict $m_\mu = m_e r = 105.66871$ MeV, and then $m_\tau =  1777.1275$ MeV from the mass sum rule (21). These are to be compared with the experimental $m_\mu = 105.65837$ MeV and $m_\tau = 1776.99^{+0.29}_{-0.26}$ MeV, showing then the errors of the order $10^{-2}\%$ both for $m_\mu $ and $m_\tau $.

I am indebted to Leszek {\L}ukaszuk for several stimulating discussions and his helpful assistance in numerical calculations.

\vfill\eject

~~~~
\vspace{0.5cm}

{\centerline{\bf References}}

\vspace{0.5cm}

{\everypar={\hangindent=0.6truecm}
\parindent=0pt\frenchspacing

{\everypar={\hangindent=0.6truecm}
\parindent=0pt\frenchspacing

[1]~Particle Data Group, {\it Review of Particle Physics, Phys.~Lett.} {\bf B 592} (2004).

\vspace{0.2cm}

[2]~For a recent discussion {\it cf. } Y.~Koide, {\tt hep--ph/0506247}, and references therein.

\vspace{0.2cm}

[3]~For a recent presentation {\it cf.} W. Kr\'{o}likowski, {\it Acta Phys. Pol.} {\bf B 33}, 2559 (2002) [{\tt hep--ph/0203107}], and references therein.

\vspace{0.2cm}

[4]~W. Kr\'{o}likowski, {\tt hep--ph/0508039}.

\vspace{0.2cm}

[5]~W. Kr\'{o}likowski, {\it Acta Phys. Pol.} {\bf B 27}, 2121 (1996); {\it Acta Phys. Pol.} {\bf B 32}, 2961 (2001) [{\tt hep--ph/0108157}]. 

\vspace{0.2cm}

\vfill\eject

\end{document}